# Query Processing Performance and Searching over Encrypted Data by using an Efficient Algorithm

Manish Sharma
M.Tech, Department of C.S
Govt. Engg. College, Ajmer
Ajmer (Rajasthan), India

Atul Chaudhary
Asst. Prof, Department of I.T
Govt. Engg. College, Ajmer
Ajmer (Rajasthan), India

Santosh Kumar
M.Tech, Department of I.T
Govt. Engg. College, Ajmer
Ajmer (Rajasthan), India

## ABSTRACT
Data is the central asset of today's dynamically operating organization and their business. This data is usually stored in database. A major consideration is applied on the security of that data from the unauthorized access and intruders. Data encryption is a strong option for security of data in database and especially in those organizations where security risks are high. But there is a potential disadvantage of performance degradation. When we apply encryption on database then we should compromise between the security and efficient query processing. The work of this paper tries to fill this gap. It allows the users to query over the encrypted column directly without decrypting all the records. It's improves the performance of the system. The proposed algorithm works well in the case of range and fuzzy match queries.

## 1. INTRODUCTION
Data is one of the most important issues for any organizations and it is very important to keep this data secure for the efficient growth of an organization. Every organization's data is stored in databases. This is a great challenge for the professionals to implement a strategy which keep the data away from the attackers. Cryptographic support is an important dimension in database security. Traditional security mechanisms cannot provide adequate security as compared to encryption mechanism for securing data. Database encryption provides the following security:

- Encryption mechanism can prevent users from obtaining data in an unauthorized manner.
- It can verify the authentic origin of a data item.
- It prevents leaking information in a database when storage mediums, like CD-ROM, tapes and disks are lost.

However, encryption degrades the system performance significantly because the Structural Query Language (SQL) queries cannot be executed directly on the encrypted data. First the encrypted data needs to be decrypt and then SQL query can be operated on it. This whole process requires a lot of processing time.

Some techniques have been proposed to get rid out from this problem of performance degradation. However, these techniques are somehow limited in their applicability. For example, hashing technique does not decrypt the content of the entire encrypted column; rather it decrypts data on the fly by using hash values of the searchable criteria [1]. However, this technique is limited to full text matches and does not work for range queries and/or fuzzy match queries.

The proposed research presents a solution which improves not only improves the performance of searching but also quickly retrieves data as compared to the existing techniques. The proposed research work presents a solution which improves not only the performance of searching but also quickly retrieves data as compared to the existing techniques. The proposed technique works with all types of queries including range queries and fuzzy match queries. It can also work with all types of encryption algorithms. This proposed algorithm is capable to retrieve only that type of records which fulfills the searching criteria, which will improve the performance of system and data confidentiality.

The rest of the paper is containing as that section II describes related work. In the section III, we discussed the actual research problem and hypothesis. In Section IV we focused on proposed security technique and section V describes how to search in the encrypted column. In Section VI we discussed about the architecture of the proposed security system and section VII presents the proposed algorithm. Section VIII gives the results of the algorithm after testing and section IX concludes the work done.

## 2. RELATED WORK
In the past time and also now the security was considered to be an extra problem in the databases. To add some extra and powerful security to data, you can use the encryption mechanism of the DBMS although encryption provides security but it has some pros and cons. Encryption reduces the system performance significantly.

[1] proposes a model for querying over encrypted data. In this proposed model they using separate chipper index for character and numeric data.

[2] proposes a hashing technique which can execute fast over encrypted data. In this work they are using hashing values as well as a number called "confuse number" which can distinguish two similar values in different records.

The Cipher index method is basically used to improve the performance and keeps the data confidentiality in the servers which is untrusted [3]. In cipher index method, the hash values are used for the searching of data on the un-trusted servers without decrypting it on the server side. This method extracts the match records to the hash values and decrypts it on the client side.

[4] uses index method for searching which could search on the range queries as well but this method is useful only for the numeric data and not useful for the character data.

In [5] a technique is proposed which can execute directly on encrypted data using a mapping function for the translation of queries from client side to server side. It preserves the data privacy and confidentiality.

## 3. RESEARCH PROBLEM AND HYPOTHESIS
The traditional search process on an encrypted column of a table is performed by decrypting the entire encrypted column and then retrieves the data which takes so much time and also reduce the performance of SELECT query in database. There are some methods which resolve the problem of performance degradation like hashing method but it is not suitable for range and fuzzy match queries.

Now we discuss our hypothesis of research:
We store our encrypted column (contain highly sensitive data) in decrypted form along with a key column which is in encrypted form in a different table. By using a separate table for searching on encrypted data resolves the problem of range and fuzzy match queries and also improves the performance of data retrieval process.

## 4. PROPOSED SCURITY TECHNIQUE

The proposed technique suggests two tables for a single main table for introducing the security in database. The first table named Encrypted_Data_Table (Table 1) contains the actual data and the second one named Query_Search_Table (Table 2) containing only that data on which the search query runs. The Encrypted_Data_Table is basically the main table of database with the only difference is that it has its sensitive column in encrypted form. The sensitive columns in the main table are encrypted using strong encryption algorithm. A copy of the sensitive data column which is in encrypted form along with the key column is taken in the "Query_Search_Table". In the Query_Search_Table, the data column is copied from the Encrypted_Data_Table is kept in the unencrypted form and the key column in the encrypted form. The order of the records in the Query_Search_Table will not be as that in the Encrypted_Data_Table. The rows of the Query_Search_Table will be reorder randomly. The encryption of the sensitive data column in the Encrypted_Data_Table and the encryption of the key column in the Query_Search_Table hide the relationship between the Encrypted_Data_Table and Query_Search_Table. The Query_Search_Table is stored in the Secure_Schema. The Secure_Schema is those schemas to which only those users are allowed who are authorized to access the encrypted data. We added Extra security by introducing some noise to those records of the Query_Search_Table from which the intruder can make some inferences.

In the proposed technique, the actual security is introduced by hiding the relationship between the Encrypted_Data_Table and Query_Search_Table. Also, the Secure_Schema makes the Query_Search_Table more secure. Similarly, in order to deceive the intruders, noise has been added to Query_Search_Table. In addition to deceive automated schema generation tools, different column headings are used for the columns of Query_Search_Table than that of the Encrypted_Data_Table.

**TABLE 1: ENCRYPTED_DATA_TABLE**

| Key | Emp_Name | Salary | Job Title |
|---|---|---|---|
| 1 | Rajesh | Encrypted | Manager |
| 2 | Suresh | Encrypted | Asst. Manager |
| 3 | Mahesh | Encrypted | Peon |
| ……. | ………… | …………….. | ………….. |

**TABLE 2: QUERY_SEARCH_TABLE**

| ABC(Key column of the Encrypted_Data_Table) | XYZ(Salary Column of the Encrypted_Data_Table) |
|---|---|
| Encrypted | 10000 |
| Encrypted | 8000 |
| Encrypted | 6000 |
| ………. | …….. |

## 5. SEARCH METHODOLOGY

Whenever any authorized user wants to search some records from the Encrypted_Data_Table and search condition is on the encrypted column, so the search will be performed on the Query_Search_Table. The search query returns keys to Encrypted_Data_Table based on the search condition, and then on the basis of that key the records will be given to the user. This technique will return only those records satisfying the user query and no extra record will be given. This result improves performance and data confidentiality.

Whenever a query is fired on the encrypted data column, the proposed algorithm performs decryption at two positions: first decryption is in the Query_Search_Table to decrypt Keys and second decryption is in Encrypted_Data_Table to decrypt actual column values. Although, it appears that this will result in more performance degradation than previous methods. The experimental results show the greater performance over the previous techniques on query over encrypted data. The reason is that the proposed approach does not need to decrypt all the values of entire encrypted column; rather it decrypts only those values which satisfy the user query. The proposed technique is very efficient whenever the amount of data retrieval is less than 40% of the total data. In the basic typical environment fewer amounts of data is retrieved in the search query, so our proposed technique is appropriate for typical environment. The searching operation of the proposed algorithm can be described well by the following example.

EXAMPLE

Reference to Table 1 Suppose a user poses the following query over the Encrypted_Data_Table.

SELECT Emp_Name, Salary
FROM Encrypted_Data_Table
WHERE Salary = 10000

The proposed algorithm interprets this query and transform as following:
SELECT Emp_Name, DecryptFunction (Salary)
FROM Encrypted_Data_Table
WHERE Key IN (SELECT DecryptFunction (ABC)
FROM Query_Search_Table WHERE XYZ =10000

Here in this query, the user wants to retrieve data of those records whose salary is 10000. The proposed algorithm performs searching on the Query_Search_Table, as the encrypted column is selected in WHERE clause of the query. The inner query performs the searching in the Query_Search_Table for the keys of those particular records which satisfies the user's search criteria. Now After this, the keys are returned in the WHERE clause of the outer query. Now, this outer query uses those keys to retrieve exactly those records which the user wants. Here, the decryption function is called twice, first for the decryptions of the keys in the inner query from Query_Search_Table and the second for the decryptions of the actual values in the outer query form the Encrypted_Data_Table.

## 6. ARCHITECTURE OF THE PROPOSED SYSTEM

The architecture for the proposed security model, depicted in figure 1, consists of three main parts. These are user,

Encrypted_Data_Table and Query_Search_Table which is stored in the "Secure Schema".

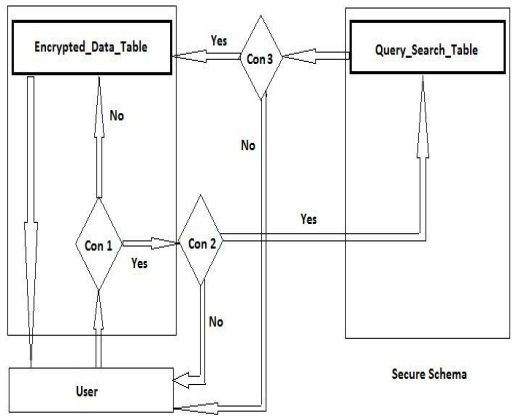

Figure 1: Architecture of the proposed security model

Con 1: Checks the condition that whether the query is on encrypted column?
Con 2: Checks the user validity to the secure schema.
Con 3: Is/are any record(s) found?

The user poses the query for searching. After that query is checked that whether it is fired on encrypted column or unencrypted column. If it is unencrypted then data is retrieved from Encrypted_Data_Table otherwise user is authenticate to enter in the "Secure Schema" and retrieval of data is performed indirectly from Encrypted_Data_Table viva Query_Search_Table.

## 7. ALGORITHMIC OUTLINES

The outlines of the proposed technique are given in the form of an algorithm shown below:

1. [User Query]
       User poses query
2. [Check the searching column]
     If (searching column is not encrypted)
       Goto step 3
     Else if (authorized user)
       Goto step 4
       Else
         Goto step 5
3. [Retrieval of data]
     Retrieve data from Encrypted_Data_Table
       Goto step 5
4. [Passing control to the Secure Schema]
     [Check for query match]
     If (no query match) then
       i. Display ("Search is unsuccessful")
       ii. Goto step 5
   Else
   [Retrieval of encrypted keys]
       i. Retrieve the corresponding encrypted key(s)
       ii. Decrypt the key(s)
       iii. Retrieve the data from Encrypted_Data_Table based on key(s).
5. Exit

## 8. TESTING AND RESULTS

The proposed algorithm is tested on a database of blood donating unit of the Govt. Hospital Jaipur, India. A table of blood donors having 59299 records was taken for testing purposes. Out of these records, data with 2% difference was retrieved up to 44% of the total data. The result of this testing is very satisfactory whenever the data retrieval percentage is less than 40% of the total. However in the typical environment there is less data retrieval so the given proposed technique is more appropriate for the typical environment.

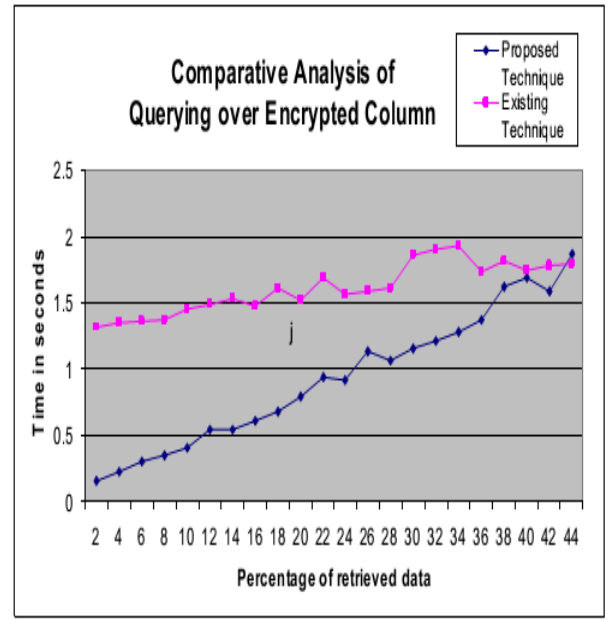

Figure 2: Analysis result of the algorithm

Here in figure 2, two curves have been shown. The upper curve, in the figure, shows the time of decryption of the total contents of entire column and its searching. While the second curve, shown bellow in the figure, represents the decryption of only those values which satisfies the user's search criteria. Our proposed technique decrypts two values for a single record retrieval. In the first decryption key value of the Query_Search_Table is decrypted and in the second the actual value of the Encrypted_Data_Table is decrypted. As per operation rules of the algorithm, both the curves should intersect at 50% of the total retrieval of the data but they intersect at 44%. The left 6% degradation of the results is due to the extra overhead of joining in between the tables: Query_Search_Table and Encrypted_Data_Table.

## 9. CONCLUSION

This research work proposed efficient algorithm for searching over encrypted data. Our proposed algorithm efficiently eliminates the limitations of the existing techniques for fuzzy match and range queries. This algorithm is efficient for searching of data whenever the retrieval of data is less than 40% of the total data.

## 10. ACKNOWLEDGEMENTS

I wish to thank Mr. Atul Chaudhary, Asst. Professor, Computer Department for his help in understanding the whole concept. I also wish to extend our heartfelt thanks to